\documentclass[aps,prb,twocolumn,showpacs,superscriptaddress]{revtex4}

\usepackage[pdftex]{graphicx} 
\usepackage{natbib}
\usepackage{verbatim}

\newcommand{\HeIa}{He~I$\alpha$}

\begin{document}

\title{Spin-orbit splitting of the Shockley surface state on Cu(111)}

\author{A. Tamai}
\altaffiliation{present address: D\'epartement de Physique de la Mati\`ere Condens\'ee, Universit\'e de Gen\`eve, 24 Quai Ernest-Ansermet, 1211 Gen\`eve 4, Switzerland}
\affiliation{SUPA, School of Physics and Astronomy, University of St Andrews, St Andrews, Fife KY16 9SS, United Kingdom}
\author{W. Meevasana}
\altaffiliation{present address: School of Physics, Suranaree University of Technology, Nakhon Ratchasima, 30000, Thailand and Thailand Center of Excellence in Physics, CHE, Bangkok, 10400, Thailand}
\affiliation{SUPA, School of Physics and Astronomy, University of St Andrews, St Andrews, Fife KY16 9SS, United Kingdom}
\author{P.D.C. King}
\altaffiliation{present address: Kavli Institute at Cornell for Nanoscale Science, Ithaca, New York 14853, USA}
\affiliation{SUPA, School of Physics and Astronomy, University of St Andrews, St Andrews, Fife KY16 9SS, United Kingdom}
\author{C. Nicholson}
\affiliation{SUPA, School of Physics and Astronomy, University of St Andrews, St Andrews, Fife KY16 9SS, United Kingdom}
\author{A. de la Torre}
\altaffiliation{present address: D\'epartement de Physique de la Mati\`ere Condens\'ee, Universit\'e de Gen\`eve, 24 Quai Ernest-Ansermet, 1211 Gen\`eve 4, Switzerland}
\affiliation{SUPA, School of Physics and Astronomy, University of St Andrews, St Andrews, Fife KY16 9SS, United Kingdom}
\author{E. Rozbicki}
\affiliation{SUPA, School of Physics and Astronomy, University of St Andrews, St Andrews, Fife KY16 9SS, United Kingdom}
\author{F. Baumberger}
\altaffiliation{present address: D\'epartement de Physique de la Mati\`ere Condens\'ee, Universit\'e de Gen\`eve, 24 Quai Ernest-Ansermet, 1211 Gen\`eve 4, Switzerland}
\affiliation{SUPA, School of Physics and Astronomy, University of St Andrews, St Andrews, Fife KY16 9SS, United Kingdom}

\begin{abstract}
We present angle-resolved photoemission data from Cu(111). Using a focused 6~eV continuous wave laser for photo-excitation, we achieve a high effective momentum resolution enabling the first detection of the Rashba spin splitting in the Shockley surface state on Cu(111). The magnitude of the spin-splitting of $\Delta k\sim 0.006$~\AA$^{-1}$ is 
surprisingly large and exceeds values predicted for the analogous surface state on Ag(111) but is reproduced by first principles calculations. We further resolve a kink in the dispersion which we attribute to electron-phonon coupling.
\end{abstract}

\maketitle

Two dimensional electron gases (2DEGs) are essential elements of electronic devices and have played a pivotal role in fundamental condensed matter physics for decades. They are most commonly realized at the interface of conventional semiconductors and have been used extensively to study the complex phases that emerge in seemingly simple electronic systems in the presence of many-body interactions, magnetic fields, disorder or lateral quantum confinement.
More recently, 2DEGs created at the interfaces and surfaces of bulk-insulating transition metal oxides \cite{nat:427:423:Ohtomo,sci:327:1607:Mannhart,arcmp:2:141:Zubko,nmat:10:114:Meevasana,nat:469:189:Santander-Syro} and topological insulators \cite{ncomm:1:128:Bianchi, prl:107:096802:King} have generated much interest because of their unconventional properties and potential for novel applications.

A different class of 2DEGs exists on the surface of many elemental metals. Whereas semiconductor 2DEGs derive from the confinement of bulk states into a thin layer, electronic surface states on metals result directly from the broken translational symmetry, which permits new, evanescent solutions of the Schr\"odinger equation in projected bulk band gaps. Prototypical examples are found at the Brillouin zone center of the noble metal (111) surfaces. These Shockley surface states derive from the gap at $L$, have a free electron dispersion, densities of a few times $10^{13}$~cm$^{-2}$ and Fermi wave lengths of 30 - 80~\AA. They played an important role in the study of lateral quantum confinement effects at metal surfaces~\cite{hel94,bur98,mug01, nat:483:306:Gomes} and attracted attention as model systems for benchmarking the capability of scanning tunneling spectroscopy and angle-resolved photoemission (ARPES) \cite{prl:82:4516:Burgi,kli00,physb:351:229:Reinert,jpcm:15:693:Reinert} in deriving intrinsic quasiparticle lifetimes in weakly correlated systems.

If confined by an asymmetric potential well breaking inversion symmetry, the spin-degeneracy in a 2DEG can be lifted by the Rashba effect~\cite{jetpl:39:78:Bychkov}. Such non-magnetic but fully spin-polarized electronic systems realized on metal surfaces~\cite{las96,prb65:033407:Nicolay,prb:69:241401:Hoesch,prl:98:186807:Ast} or in semiconductor hosts ~\cite{prl:78:1335:Nitta,prl:107:096802:King, ncomm:3:1159:King} have been investigated intensely 
 for fundamental reasons and because of their importance for spintronic devices, most notably the spin field-effect transistor.~\cite{apl:56:665:Datta,sci:325:1515:Koo}
In this context, the Shockley surface state on Au(111) served as an early model system permitting the first direct spectroscopic measurements of the characteristic spin-momentum locking in a Rashba split 2DEG.~\cite{las96,prb65:033407:Nicolay,prb:69:241401:Hoesch} However, despite the large number of spectroscopic studies of the analogous surface states on the (111) surfaces of the lighter noble metals, a Rashba splitting could not be observed to date for Cu and Ag and is thus generally assumed to be below the resolution of electron spectroscopic techniques.

Here we present new ARPES data from Cu(111). Using a 6~eV laser focused to a spot size of $\sim3$~$\mu$m
as excitation source, we successfully reduce the common broadening of photoemission lines due to the integration over residual inhomogeneity and roughness of the sample surface and simultaneously increase the momentum resolution of the electron optics. Our experimental line widths at the Fermi level are more than a factor of three narrower than in the best published data. This is sufficient to resolve for the first time a clear Rashba-type splitting in the dispersion of the Shockley surface state.

A clean Cu(111) surface was prepared by repeated cycles of Ar ion sputtering followed by annealing at 500$^{\circ}$~C. Photoelectrons were excited using a continuous wave (cw) laser system consisting of a 820~nm diode laser with tapered amplifier (Sacher Lasertechnik) and two frequency doublers with actively stabilized bow-tie cavities (LEOS solutions). This setup provides $>1$~mW cw radiation at a wavelength of 205~nm ($\sim 1\cdot10^{15}$~photons/s with $h\nu=6.05$~eV) in a neV band width and a stability of the photon energy of better than 10~$\mu$eV during a typical experimental run of a few hours. For the experiments presented here, the UV radiation was focused into a spot of $\sim3$~$\mu$m diameter on the sample and the photon flux was reduced to around $10^{13}$~photons/s in order to protect the detector from excessive count rates. Control experiments were performed with a monochromatized He plasma light source. Photoelectrons were analyzed using a SPECS Phoibos 225 spectrometer. Its energy and angle resolution was set to $<2.5$~meV / $\sim0.3^{\circ}$ and all measurements were done at 6~K. Density functional calculations (DFT) including spin-orbit coupling were performed using the Wien2k code~\cite{bla2k} for bulk truncated slabs with thicknesses up to 31 layers of Cu.

\begin{figure*}[htb]
\includegraphics[width=17.5cm]{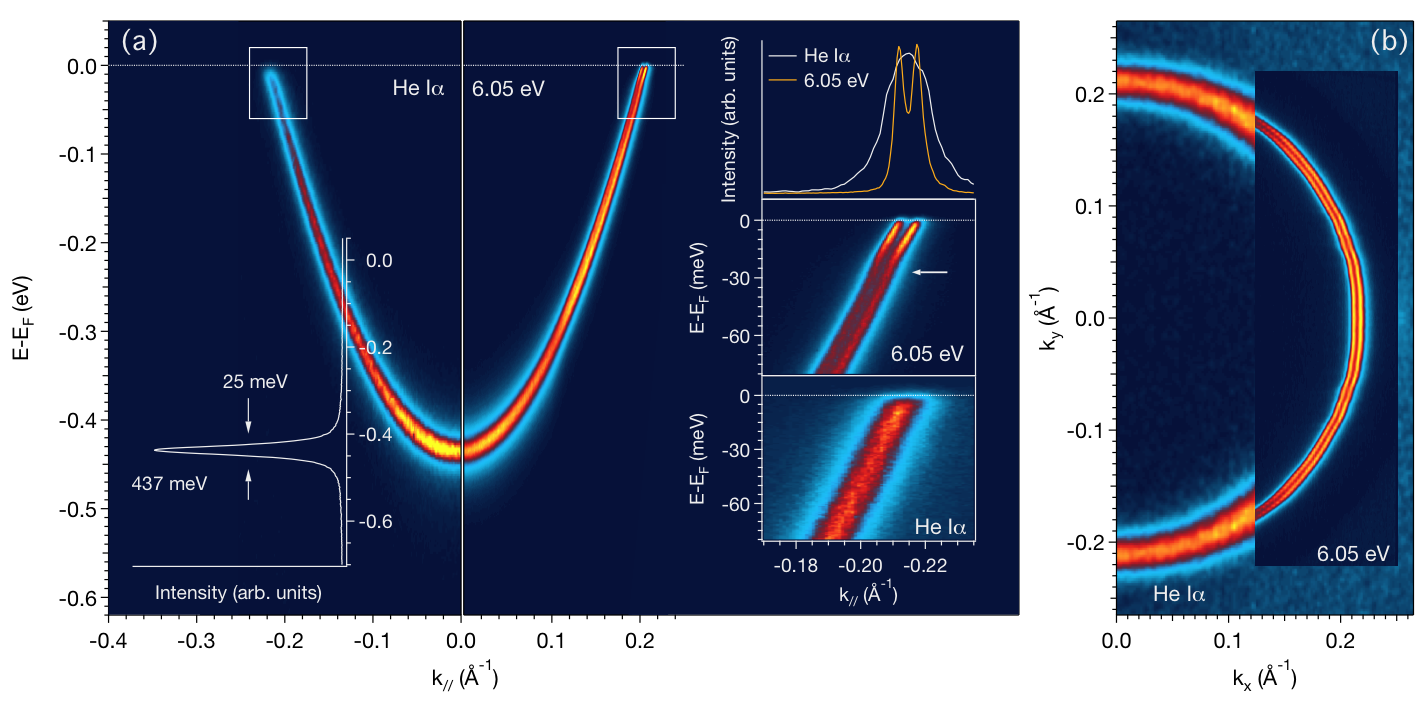}
\caption{Conventional and laser-ARPES data from the Cu(111) surface state. (a) shows the full parabolic dispersion measured with He I$\alpha$ (left) and laser excitation (right). The insets show momentum distribution curves at the Fermi level (E$_{F}$) and expand the most crucial region of the dispersion near E$_{F}$ revealing the momentum independent splitting of the dispersion, characteristic of Rashba systems with small wave vectors. (b) shows a section of the Fermi surface measured with He~I$\alpha$ and laser excitation respectively.}
\end{figure*}

The Shockley surface state on Cu(111) is an iconic electronic system and has been investigated by angle resolved photoemission for more than 30 years.~\cite{gar75,kev87,rei01} We therefore start the discussion of our new observations by demonstrating that the laser-ARPES data is fully consistent with measurements at higher photon energy. This is non-trivial since the sudden approximation underlying the interpretation of conventional ARPES data in terms of the single particle spectral function ultimately must fail as the photon energy approaches the ionization threshold. Eventually, the interaction of the photoelectron with the $(N-1)$ system left behind will become non-negligible and increasing relaxation of the latter might reduce the photoemission line width and induce a shift towards higher kinetic energy. In order to check for such signatures of a failure of the sudden approximation we compare in Fig.~1 laser-ARPES data with measurements from the same sample performed with a monochromatized He discharge source using the \HeIa\ line with $h\nu=21.2$~eV. The gross dispersion measured with laser-excitation can be fitted with an effective mass of $\sim$~0.41~m$_e$ and a binding energy of $\sim$~437~meV, in good agreement with our \HeIa\ data and the literature.~\cite{gar75,kev87,rei01} The quantitative agreement extends to the line width at the band bottom. From a fit to a Lorentzian convolved with the (nearly negligible) instrumental resolution we find a full width at half maximum of 25~meV, in excellent agreement with the best published conventional ARPES data from Cu(111) \cite{rei01} and theoretical results.~\cite{kli00,prl:88:066805:Eiguren} This strongly suggests that all many-body interactions contributing to the quasiparticle lifetime are contained in the laser-ARPES spectra.

The splitting of the quasiparticle band is barely visible on the scale of the full occupied band width in Fig.~1 but revealed clearly in the expanded view of the near-E$_F$ region in the laser-ARPES data. This panel further reveals a subtle kink in the dispersion of both branches around an energy of 30~meV which we attribute to weak electron-phonon coupling. Although this kink is hardly discernible in the \HeIa\ data, its position and magnitude are consistent with previous measurements at higher photon energy that detected a very small reduction of the line width near E$_F$.~\cite{prl:88:066805:Eiguren} This confirms that low-energy loss features including signatures of electron-phonon coupling in ARPES data taken with photon energies around 6~eV can be interpreted within the sudden approximation, consistent with earlier findings on cuprates.~\cite{prl:96:017005:Koralek} Finally, we show in the top-right inset in Fig.~1(a) that the observed splitting is smaller than the line width of a resolution limited momentum distribution curve taken with \HeIa\ radiation. Hence, our laser-ARPES data is fully consistent with conventional ARPES measurements of the Cu(111) surface. The qualitative differences to earlier work can thus be attributed to instrumental advances leading to a significantly improved resolution.

\begin{figure}[htb]
\includegraphics[width=7.5cm]{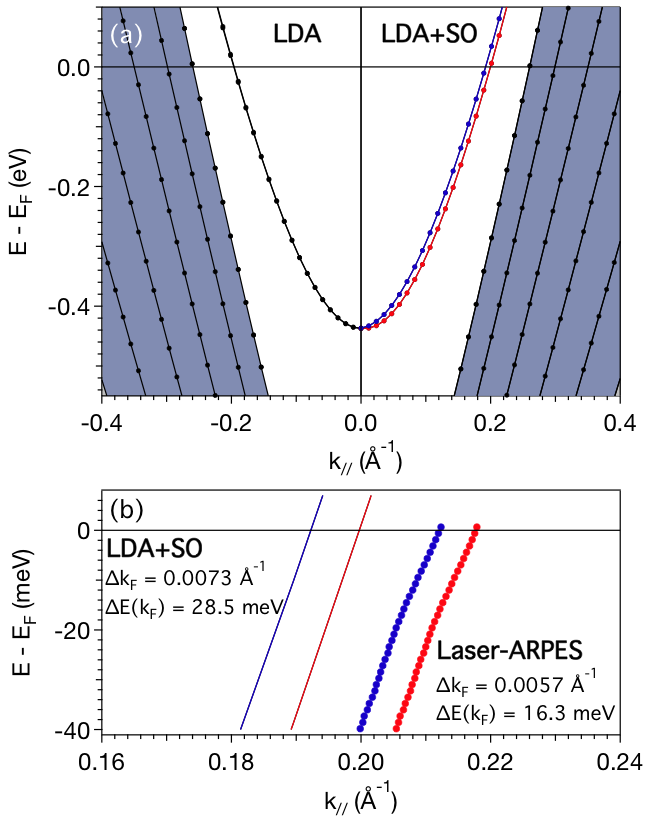}
\caption{(a) Density functional calculations of the surface state dispersion in a slab of 31 layers. The inclusion of spin-orbit interaction results in a Rashba-type splitting of the dispersion consistent with the experiments. (b) Quantitative comparison between the LDA+SO calculation and the laser-ARPES data.}
\end{figure}

Fits to momentum distribution curves indicate that within our experimental resolution the splitting $\Delta k$ in the dispersion does not depend on binding energy and Fermi surface angle. The dispersion is therefore described by two parabolic subbands of equal effective mass and energy at the band bottom, which is characteristic for a Rashba splitting of a free electron like band near the Brillouin zone center. However, the magnitude of the splitting of $\Delta k\sim 0.0057$~\AA$^{-1}$ is surprising. Contrary to naive expectations based on atomic spin-orbit splittings, we find that $\Delta k$ is significantly larger than calculated for Ag(111) and only four times smaller than found experimentally and theoretically for Au(111).~\cite{las96,prb65:033407:Nicolay,prb:69:241401:Hoesch}
It was, however, pointed out by several authors that such simplistic scaling relations fail to quantitatively explain existing data and density functional calculations of the band structure for a number of spin-orbit coupled systems.~\cite{jpcm:15:693:Reinert,ss:600:3888:Bihlmayer} Rather, the magnitude and also the sign of the Rashba parameter appear to be dominated by the asymmetry of the wave function near the nuclei, which can be obtained from electronic structure calculations.~\cite{ss:600:3888:Bihlmayer,prb:84:115426:Bentmann} 
Indeed, the experimentally observed splitting on Cu(111) is in fair agreement with our density functional calculations that include spin-orbit interaction, as shown in Fig.~2. This strongly supports the interpretation of our data in terms of a Rashba splitting of the Shockley surface state.
%

\begin{figure}[htb]
\includegraphics[width=7.5cm]{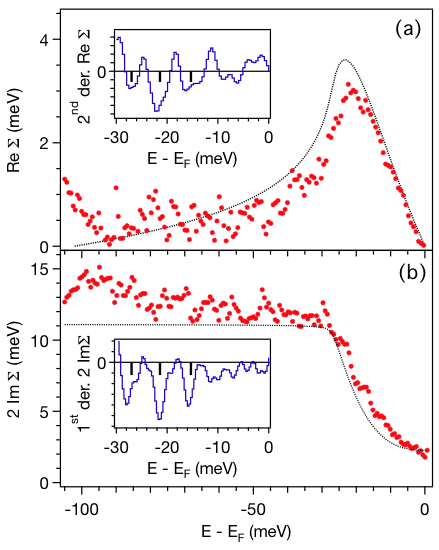}
\caption{(a,b) Real and imaginary part of the self-energy of the Shockley surface state obtained from fits to the laser-ARPES data. A Debye model of the electron-phonon interaction calculated for T=6~K and a coupling constant $\lambda=0.16$ is shown as a thin dotted line. The fine structure in the self-energy is visualized through the second derivative of Re$\Sigma$ and the first derivative of Im$\Sigma$, respectively, shown in the insets to (a,b). The minima of this function at 15, 21, 27~meV indicate the frequency of maxima in the Eliashberg function $\alpha^2F(\omega)$ and are shown as vertical bars.}
\end{figure}
We now focus on the quasiparticle self-energy near the chemical potential extracted from our laser-ARPES data with improved resolution. The raw dispersion data displayed in Fig.~1 shows a clearly discernible kink at an energy of $\sim 30$~meV (white arrow), indicating a sharp structure in the self-energy $\Sigma$ at this frequency. In order to extract quantitative values of $\Sigma$ we approximate the line shape of the spin-split momentum distribution curves by two Lorentzians with a width of 2~Im$\Sigma$ convoluted with a Gaussian accounting for the experimental resolution. The real part of the self-energy is obtained by subtracting a parabolic fit of the bare electron dispersion from the energy of the quasiparticle pole. The gross shape of the real and imaginary part of $\Sigma$ follows a simple Debye model for the electron-phonon-interaction with an Eliashberg function $\alpha^2F(\omega)\propto\omega^2$ for frequencies up to $\omega_{\rm{Debye}}=27$~meV and a coupling constant of $\lambda=0.16$ (dotted line in Fig.~3(a,b)), in good agreement with the value deduced in Ref.~\cite{mcd95} from the temperature dependence of the line width.
Assuming a momentum resolution of $\sim 0.003$~\AA$^{-1}$ / $0.3^{\circ}$ full width at half maximum we find a residual quasiparticle line width 2~Im$\Sigma$ at the chemical potential of $\sim 2$~meV, consistent with a very high surface quality. The total phonon contribution to the line width, given by the increase of 2~Im$\Sigma$ between $\omega=0$ and $\omega_{\rm{Debye}}$ is $\sim 10$~meV, significantly larger than found in the many-body calculations by Eiguren \textit{et al.}~\cite{prl:88:066805:Eiguren} but in excellent agreement with the experimental data of the same authors taken with He~I radiation.~\cite{physb:351:229:Reinert,prl:88:066805:Eiguren}

Intriguingly, our data show a subtle fine structure in the self energy in the form of faint shoulders in the real part and small steps in the imaginary part at corresponding energies. This is indicative of a more complex Eliashberg function $\alpha^2F(\omega)$ with multiple maxima.~\cite{prl:95:117001:Zhou} In order to reveal their position, we show the second derivative of Re$\Sigma$ and the first derivative of Im$\Sigma$ in the insets of Fig.~3(a,b), respectively. The minima of these functions correspond to peaks in $\alpha^2F(\omega)$ and are observed most clearly at frequencies of 15, 21 and 27~meV (marked by black bars), in excellent agreement with the dominant contributions from bulk phonon modes to $\alpha^2F(\omega)$ calculated by Eiguren \textit{et al.}.~\cite{prl:88:066805:Eiguren} At very low energy the results from our analysis are less conclusive. However, rather than the single sharp surface phonon mode at 13~meV predicted by Eiguren \textit{et al.}, our data indicate several weak peaks of $\alpha^2F(\omega)$ in the range of 7 - 11~meV.
We note that resolving these features approaches the limits of the imaging precision of modern electron spectrometers. In order to minimize artifacts introduced by the electron optics, we performed these experiments without grid in front of the multi-channel plate of the detector and selected a geometry that images the relevant energy range below the Debye frequency onto the most homogeneous part of the entrance slit to the hemisphere.

We finally comment on the surface sensitivity of laser-ARPES. The universal curve of the electron inelastic mean free path \cite{sea79} suggests a probing depth around 40~\AA{} for $h\nu\sim 6$~eV, comparable to excitation with hard x-rays. Yet, the intensity from the Cu surface state is large and exceeds that of bulk states in several layered materials that we investigated with the same setup. We attribute this to a particularly favorable matrix element. The Shockley surface state is located in the projected gap at $L$ and its transition matrix element peaks for the same perpendicular final state momenta, which can be reached with a photon energy around 70~eV~\cite{kev86} and are again approached at very low photon energies, as used in this study and by Gartland \textit{et al.}.~\cite{gar75} Moreover, for a surface state on a 3D free electron metal band-like final states are available for most excitation energies because they strongly disperse in $k_{z}$. This is generally not the case for layered materials where band-like final states disperse weakly in $k_{z}$ and are thus available for selected photon energies only, which frequently suppresses matrix elements at low photon energies in quasi-2D materials.

In conclusion we presented new laser-ARPES measurements from Cu(111) that resolve for the first time a Rashba-type spin splitting in the Shockley surface state. This demonstrates that instrumental advances in ARPES continue to reveal unexpected effects, promising new insight even in the most widely investigated systems.

We gratefully acknowledge discussions with 
M. Grioni, J. Osterwalder and E. Rotenberg.
This work has been supported by the European Research Council, the Scottish Funding Council and the UK EPSRC. Computations were carried out on the EaStCHEM Research Computing Facility.


\end{document}